\documentclass[12pt]{iopart}

\usepackage{graphicx,refs}

\newcommand{\btosg}{\ensuremath{b \to s g}}
\newcommand{\btosgg}{\ensuremath{b \to s gg}}
\newcommand{\btosff}{\ensuremath{b \to s \bar f f}}

\newcommand{\mc}{\ensuremath{m_c}}
\newcommand{\mb}{\ensuremath{m_b}}

\newcommand{\BR}{{\ensuremath{\mathcal{B}}}}
\newcommand{\BRbar}{{\ensuremath{\overline{\BR}}}}

\newcommand{\BRc}{\ensuremath{\BR_{c\!\! /}}}
\newcommand{\BRsl}{\ensuremath{\BR_{\rm sl}}}
\newcommand{\BRbarC}{\ensuremath{\BRbar_{c\!\! /}}}

\newcommand{\BRbarNC}{\ensuremath{\overline{\BR}(B \to {\rm no\ charm})}} 

\newcommand{\nc}{\ensuremath{n_c}}
\newcommand{\VubVcb}{\ensuremath{|V_{ub}/V_{cb}|}}

\newcommand{\BRfi}[2]{\ensuremath{\BR(#1 \to #2)}}

\newcommand{\LQCD}{\ensuremath{\Lambda_{\rm QCD}}}

\newcommand{\tento}[1]{\cdot 10^{#1}}

\newcommand{\abstracttext}{
  I present the Standard Model calculation of the decay rate $\Gamma(\btosg)$
  ($g$ denotes a gluon) at next-to-leading logarithms (NLL). In order to get a
  meaningful physical result, the decay \btosgg\ and certain contributions of
  \btosff\ (where $f$ are the light quark flavours $u$, $d$ and $s$) have to
  be included as well.  Numerically we get $\BR^{\rm NLL}(\btosg) = (5.0 \pm
  1.0) \tento{-3}$ which is more than a factor 2 larger than the leading
  logarithmic result $\BR^{\rm LL}(\btosg) = (2.2 \pm 0.8)\tento{-3}$. Further,
  I consider the impact of this contribution on the charmless hadronic
  branching ratio \BRc, which could be used to extract the CKM-ratio \VubVcb\ 
  with more accuracy. Finally, I have a short look at \BRc\ in scenarios where
  the Wilson coefficient $C_8$ is enhanced by new physics.
}

\begin{document}


\begin{titlepage}
  
  \begin{flushright}
    BUTP--00/32    \\
    hep-ph/0011093      \\[3ex] 
  \end{flushright}
  \vspace{2.5cm}
  
  \begin{center}
    {\bf The decay $\bi{\btosg}$ at NLL in the Standard Model}\\[5ex] 
    P.~Liniger\\[1ex]
    {\it Institut f\"ur theoretische Physik, 
      Universit\"at Bern}, \\  
    {\it Sidlerstrasse 5, 3012 Bern, Switzerland} \\[10ex]
  \end{center}
  \begin{center} 
    ABSTRACT \\
    \vspace*{1mm}
    \parbox{13cm}{
      \abstracttext
      }
  \end{center}
  
  \vspace*{2truecm}
  {\begin{center} 
      {\it Talk presented at the}\\ 
      {\it ``UK Phenomenology Workshop on   Heavy Flavour and CP Violation''}      \\
      {\it St John's College, Durham, 17 - 22 September 2000}
    \end{center}}
  \vfill
\end{titlepage}

\thispagestyle{empty}


\title[The decay \btosg\ at NLL in the Standard Model]{The decay $\bi{\btosg}$ at NLL in the Standard Model}

\author{Patrick Liniger\footnote{Work done in collaboration with Christoph
    Greub \cite{cgpl}. Work partially supported by Schweizerischer
    Nationalfonds.}}
 
\address{Institut f\"ur theoretische Physik, Universit\"at Bern, CH-3012 Bern}

\ead{liniger@itp.unibe.ch}

\begin{abstract}
  
  \abstracttext

\end{abstract}

\section{Introduction}
\label{sec:intro}

Theoretical studies for inclusive $B$-decays have become accessible due to
heavy quark effective theory (HQET). HQET \cite{Bigi1,Bigi2} states that the
amplitudes of decaying $B$-mesons can be expanded in powers of $(\LQCD/\mb)$
where the leading term is nothing but the decay of the underlying $b$-quark.
Corrections to this start at $\Or(\LQCD^2/\mb^2)$ only, which numerically
amounts to $\sim 5\%$.

One interesting subclass of inclusive $B$-decays are the charmless ones $B \to
X_{c\!\!/}$. They have been thoroughly investigated in the literature: the
calculation for $b\to \bar q' q' q$ with $q=d,s$ and $q'=u,d,s$ have been
available to NLL for quite some time \cite{Altarelli,Lenz1,Lenz2,Lenz3} and
the missing piece, \btosg, is what is presented here. Due to the sensitivity
of the charmless branching ratio \BRc\ to the poorly known CKM ratio \VubVcb,
these decays might be used in order to get a better determination of this
standard model parameter.

The decay \btosg\ turned out to be an interesting issue in the discussion of
the 'missing charm puzzle'; for a long time there was a discrepancy between
theory and experiment for the average charm multiplicity \nc\ per $B$-decay
and for the inclusive semileptonic branching ratio \BRsl\ \cite{Bigi_Falk}.
This issue seems to have settled a bit.

\section{Theoretical Framework}
\label{sec:theo}

The calculation is based on an effective Hamiltonian $\mathcal{H}_{\rm eff}$
obtained by integrating out the heavy degrees of freedom of the standard model
(i.e. the $t$-quark and the $W$-boson). Retaining operators up to dimension
six, the 5-flavour effective Hamiltonian responsible for \btosg\ reads
\begin{eqnarray}
  \label{eq:heff}
  \mathcal{H}_{\rm eff} = - \frac{4 G_F}{\sqrt2} \,V_{ts}^* V_{tb} 
   \sum_{i=1}^8 C_i(\mu) O_i(\mu),
\end{eqnarray}
where $V_{ij}$ are CKM-matrix entries, $C_i(\mu)$ the Wilson coefficients and
$O_i(\mu)$ the relevant operators. As the Wilson coefficients $C_{3-6}$ are
small, we only consider the operators
\begin{eqnarray}
  \label{eq:operators}
  O_1 \,= 
  (\bar{s}_L \gamma_\mu T^A c_L)\, 
  (\bar{c}_L \gamma^\mu T^A b_L)\,, &
  O_2 \,=  (\bar{s}_L \gamma_\mu c_L)\, 
  (\bar{c}_L \gamma^\mu b_L)\,, \nonumber \\
  O_8 \,=
  \frac{g_s}{16\pi^2} \,{\bar m}_b(\mu) \,
  (\bar{s}_L \sigma^{\mu\nu} T^A b_R)
     \, G^A_{\mu\nu} \ . & 
\end{eqnarray}
Here $T^A$ stand for the $SU(3)_{\rm{colour}}$ generators. The small CKM
matrix element $V_{ub}$ as well as the $s$-quark mass are neglected. The whole
work is done in the NDR scheme, i.~e. with anticommuting $\gamma_5$ and using
$\overline{\rm MS}$ subtraction.

\section{Contributions}
\label{sec:contrib}

\subsection{Virtual Corrections to $O_1$ and $O_2$}
\label{sec:O12}

\begin{figure}[htbp]
  \begin{center}
    \caption{Complete list of two-loop Feynman diagrams for $b \to s g$
      associated with the operators $O_1$ and $O_2$. The crosses denote the
      possible locations where the gluon is emitted.}
    \label{fig:O12}
    \hfill \includegraphics[width=13cm]{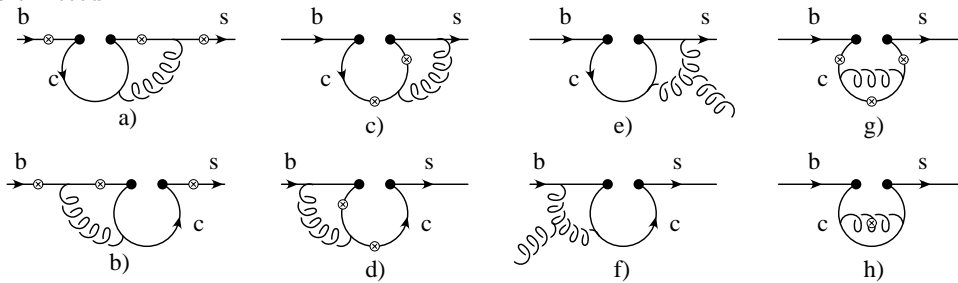}
  \end{center}
\end{figure}

In \fref{fig:O12} there is the complete set of Feynman diagrams that
contribute to \btosg. As these start at $\Or(g_s\alpha_s)$ only (no one loop
contributions), the Wilson coefficients $C_1$ and $C_2$ are needed to leading
logarithmic precision only.

The heart of the procedure is to make a Mellin-Barnes representation
\cite{Mellin} of the two-loop integrals. That is, after the Feynman
parametrization and after integrating over the loop momenta, the remaining
expression is of the form $(k^2 - M^2)^{-\lambda}$ which can be written as
\begin{eqnarray}
  \label{eq:Mellin}
      \frac{1}{(k^2 - M^2)^\lambda} = \frac{1}{(k^2)^\lambda} \,
    \frac{1}{\Gamma(\lambda)} \, \frac{1}{2 \pi \rmi} \, \int_{\gamma} \rmd{s}
    (-M^2/k^2)^s \Gamma(-s) \Gamma(\lambda+s)\;,
\end{eqnarray}
where, in the actual calculation, $M^2$ (resp. $k^2$) is $\mc^2$ (resp.
$\mb^2$) times some combination of the Feynman parameters and $\gamma$ is a
path in the $s$-plane parallel to the imaginary axis with $-\lambda < \Re(s) <
0$. This representation leads naturally to an expansion in $z=\mc^2/\mb^2$
which is numerically around $0.1$. All contributions up to $z^3$ are
retained. In the end, the result consists of powers and logarithms of $z$ only.

The matrix elements for the operators $O_{1/2}$ contain UV-divergences which
are removed by renormalization; there are no IR problems and also
singularities due to $m_s=0$ are absent.

\subsection{Virtual Corrections to $O_8$}
\label{sec:O8}

\begin{figure}[htbp]
  \begin{center}
    \leavevmode
    \caption[f1]{Complete list of Feynman diagrams associated with the 
      operator $O_8$. The analogues of a), c) and d), where the
      virtual gluon hits the $b$-quark instead of the $s$-quark,
      are not shown explicitly.
      The real gluon can
      be attached to any of the crosses shown in a).}
    \hfill \includegraphics[width=13cm]{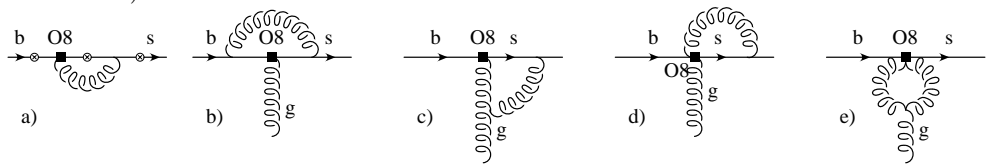}
    \label{fig:O8}
  \end{center}
\end{figure}

Some comments on the contributions of $O_8$ are in order. The relevant diagrams
are displayed in \fref{fig:O8}.

Some technical details of the calculation: for this part, the $s$-quark mass
has not been set to zero but was kept as a regulator in order to distinguish
between mass-singularities and IR-singularities.

As $O_8$ contributes to \btosg\ at tree-level already, the Wilson coefficient
$C_8$ is needed at NLL precision \cite{Misiak97}. Also due to the appearance
of the gluon field strength in the operator, $Z_3$ enters upon
renormalization. From the gluon self-energy there are logarithms
$\ln(\mu/m_f)$ (with $f=u,d,s$) entering the calculation. In order to avoid
the singularities that would show up when the respective masses are set to
zero, also the decay \btosff\ (arising from $O_8$ only) was included.

After renormalization $\Gamma_8(\btosg + s \bar f f)$ has the form
\begin{eqnarray}
  \label{eq:O8}
\Gamma_8(\btosg + s \bar f f)= \sum_{\begin{array}{l}\scriptstyle
  n,m=0,1,2\\[-1ex] \scriptstyle n+m \leq 2\end{array}}
  a_{nm}\frac{\ln^n(m_s/\mb)}{\epsilon_{\rm IR}^m}
\end{eqnarray}
where it is anticipated that the remaining singularities will cancel against
the gluon brems\-strahlung.

\subsection{Bremsstrahlung \btosgg}
\label{sec:brems}

In order to get a physically meaningful (in particular an IR finite) result,
the bremsstrahlung contributions form $O_{1/2/8}$ have to be added. Again (for
$O_8$ only, as the ones from $O_{1/2}$ are finite anyway) $m_s$ is kept as a
regulator and the phase space integrals are done in $d=4-2\epsilon$
dimensions.  Only the singularities are worked out analytically, the phase
space for the finite parts is done numerically.

\section{Result}
\label{sec:result}

Adding the various pieces, the singularities cancel in fact (according to the
KLN-theorem). The result can be cast into a form involving an effective matrix
element $\bar D$ (where $\Gamma^{\rm NLL}(\btosg)$ denotes  $\Gamma(\btosg(g)) +
\Gamma_8(\btosff)$):
\begin{eqnarray}
  \Gamma^{\rm NLL}(\btosg) =
  \frac{\alpha_s m_b^5}{24 \pi^4} G_F^2 |V_{ts}^* V_{tb}|^2 |\bar D|^2 +
  \Gamma^{\rm brems}_{\rm fin}
\end{eqnarray}
with
\begin{eqnarray}
  \label{eq:DBar}
  \bar D = &  C_8^{0,\rm{eff}} + 
  \frac{\alpha_s}{4\pi} \, \Bigl\{
  C_8^{1,\rm{eff}} 
  - \case{16}{3} C_8^{0,\rm{eff}} + C_1^0 [\ell_1 \, \ln(\mb/\mu) +
  r_1]  \nonumber \\
  &
  + C_2^0 [\ell_2 \, \ln(\mb/\mu) +r_2] + C_8^{0,\rm{eff}}
  [(\ell_8+8+\beta_0) \, \ln(\mb/\mu) + r_8] \Bigr\}
\end{eqnarray}
where $\Gamma^{\rm brems}_{\rm fin}$ denotes the finite bremsstrahlung part
which affects the result by $\sim 5\%$ only; $\ell_{1/2}$ and
$(\ell_8+8+\beta_0)$ are nothing but the entries of the anomalous dimension
matrix $\gamma^{0,\rm eff}_{18/28/88}$ and $r_{1/2/8}$ encode the loop
functions. For the Wilson coefficients we have used the perturbative
decomposition $C_i = C_i^0 + \frac{\alpha_s}{4\pi} C_i^1$.

The branching ratio is given by the expression $\BRfi{b}{sg} =
\frac{\Gamma^{\rm NLL}(\btosg)}{\Gamma_{\rm sl}} \BR_{\rm sl}^{\rm exp}.$
\begin{figure}[htbp]
  \begin{center}   
    \leavevmode
    \caption{Scale dependence of the function $\bar D$ (equation
      \eref{eq:DBar}) and the branching ratio in various approximations: the
      dashed line corresponds to leading order (for $\bar D$ this is simply
      $C_8^{0,\rm eff}(\mu)$), the solid line is the full NLL result. For the
      dotted line we set $r_1=r_2=r_8=0$ (and, for the branching ratio,
      $\Gamma_{\rm brems}^{\rm fin}=0$ in addition). The dot-dashed line is
      obtained by putting $r_2=0$ only.}  
    \vspace{0.5cm} 
    \hfill
    \includegraphics[width=6cm]{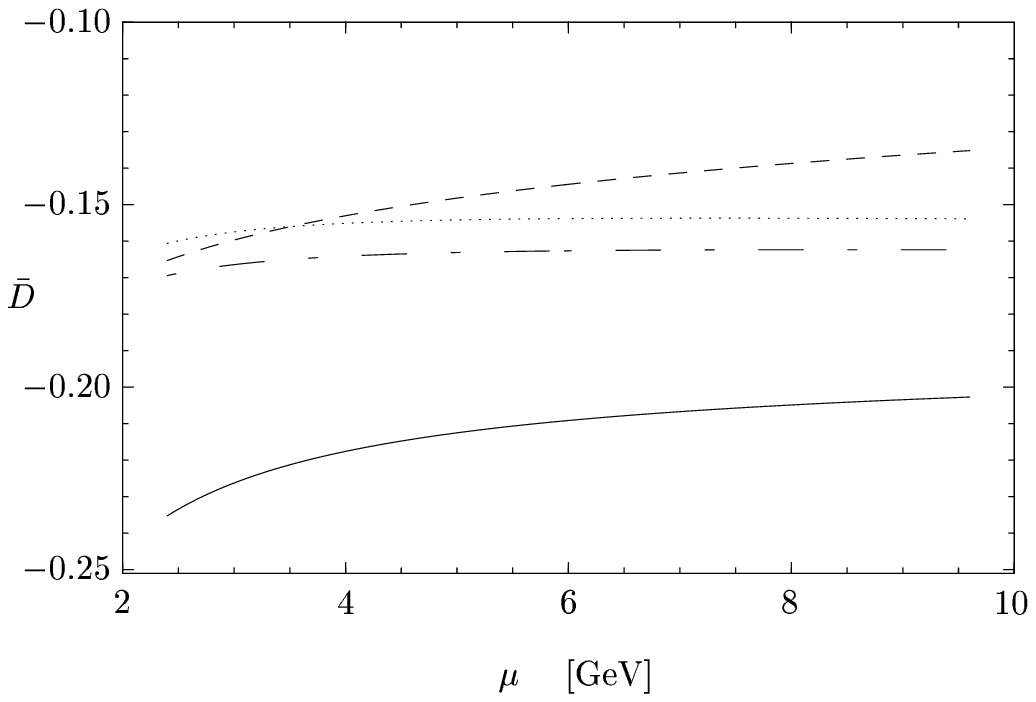} 
    \hspace{1cm}
    \includegraphics[width=6cm]{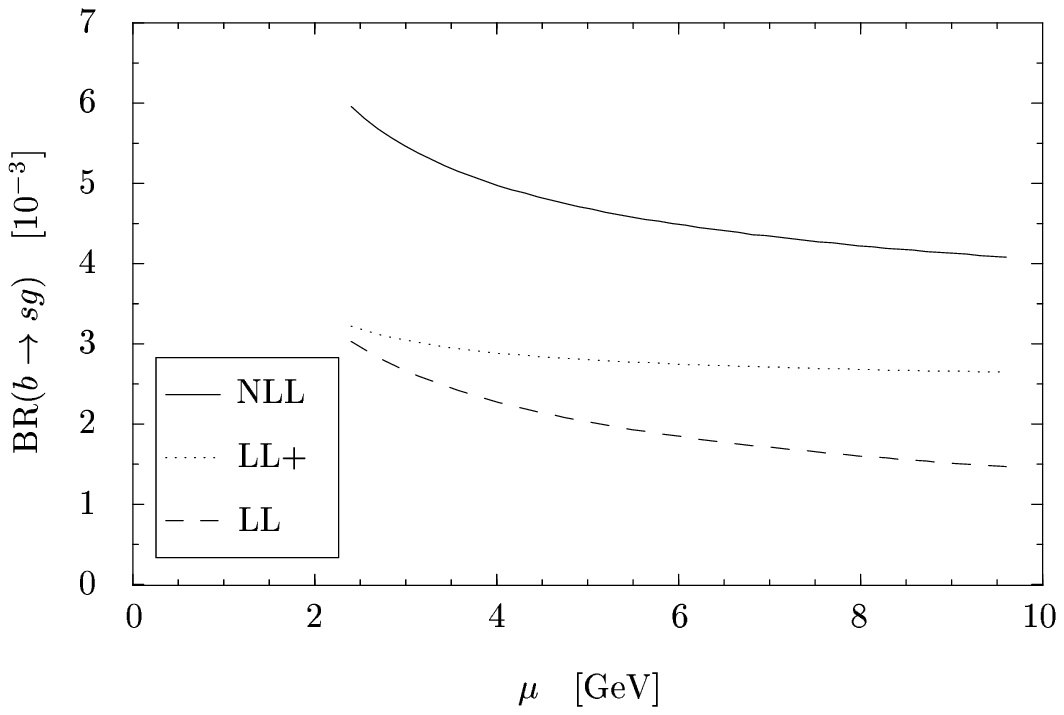}
    \label{fig:muDep}
  \end{center}
\end{figure}

\noindent
The NLL branching ratio still suffers from a big $\mu$-dependence. As is
illustrated in \fref{fig:muDep} this is due to $r_2$ which is also responsible
for the sizable enhancement of the branching ratio. This large factor is
multiplied by $\alpha_s(\mu)C_2(\mu)$ and numerically spoils the cancellation
of the $\mu$ dependence which is established through $\ell_{1/2/8}$.

Numerically we get (and reproduce the LL result \cite{Ciuchini})
\begin{eqnarray}
  \label{eq:numerix} \fl
  \BR^{\rm NLL}(\btosg) = (5.0 \pm 1.0)\tento{-3},
  \qquad & 
  \BR^{\rm LL}(\btosg) = (2.2 \pm 0.8)\tento{-3}.
\end{eqnarray}

Combining our NLL result with the work of Lenz et~al. \cite{Lenz1,Lenz2}, we
calculate the CP-averaged {\em charmless hadronic} branching ratio \BRbarC\ 
and the CP-averaged {\em total charmless} branching ratio \BRbarNC\ (the
errors are estimated by varying the scale $\mu$ in the range $[\mb/2,2\mb]$)
\begin{eqnarray}
  \label{eq:charmlessHadronic}
  \BRbarC =  (1.88^{+0.60}_{-0.38})\%, \qquad &
  \BRbarNC =  (2.22^{+0.60}_{-0.38})\%.
\end{eqnarray}
The latter is very sensitive to \VubVcb: varying this ratio in the range
$0.06-0.13$ the total charmless branching ratio covers the range
$(1.50-2.42)\%$.

\section{Remarks on $\bi{\nc}$ and $\bi{{\cal B}_{\rm\bf sl}}$}

\begin{figure}[htbp]
  \begin{center}
    \caption{Charmless hadronic branching ratio \BRbarC\
      as a function of the ratio
      $f=C_8^{\rm{eff}}(m_W)/C_8^{\rm{eff,SM}}(m_W)$. The dotted (solid) curve
      includes the LL (NLL) approximation for \btosg.}
    \includegraphics[width=8cm]{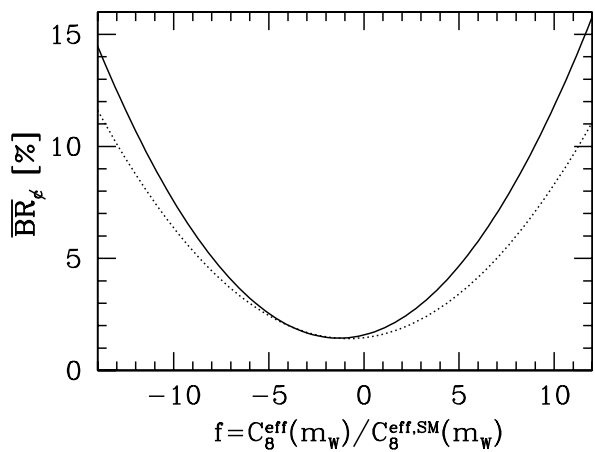}
    \label{fig:C8enh}
  \end{center}
\end{figure}

As discussed in the introduction, the theoretical predictions for both, the
charm multiplicity \nc\ and the semileptonic branching ratio \BRsl\ used to be
in disagreement with the experimental data \cite{Lenz3}. This discrepancy
decreased a lot after the inclusion of the complete NLL corrections to $b\to
c\bar u q$ and $b\to c\bar c q$ ($q=d,s$) \cite{Bagan}. If one allows the
renormalization scale $\mu$ to be as low as $\mb/4$, then there is a marginal
overlap between theory and CLEO- and LEP data \cite{Japantalk}. It is
therefore a matter of taste if one considers this problem to be solved or if
one is inclined towards an enhancement of the Wilson coefficient through non
standard model physics.  The impact of such an enhancement is displayed in
\fref{fig:C8enh}.

\section*{References}
\begin{thebibliography}{100}


\bibitem{cgpl}
C.~Greub and P.~Liniger,
hep-ph/0009144;
%
hep-ph/0008071.


\bibitem{Bigi1}
I. Bigi et al., Phys. Rev. Lett.~{\bf 71}, 496 (1993);\\
A. Manohar and M.B. Wise, Phys. Rev.~{\bf D49}, 1310 (1994);\\
B. Blok et al., Phys. Rev.~{\bf D49}, 3356 (1994);\\
T. Mannel, Nucl. Phys.~{\bf B413}, 396 (1994);\\
A. Falk, M. Luke, and M. Savage, Phys. Rev.~{\bf D49}, 3367 (1994).

\bibitem{Bigi2}
I. Bigi et al.,
Phys. Lett.~{\bf B293}, 430 (1992); 297 (1993) 477 (E).

\bibitem{Altarelli}
 G. Altarelli and S. Petrarca, Phys. Lett.~{\bf B261}, 303 (1991).

\bibitem{Lenz1}
 A. Lenz, U. Nierste and G. Ostermaier,  Phys. Rev.~{\bf D56}, 7228 (1997).

\bibitem{Lenz2}
 A. Lenz, U. Nierste and G. Ostermaier,  Phys. Rev.~{\bf D59}, 034008 (1999).


\bibitem{Lenz3}
 A.~Lenz, these proceedings.


\bibitem{Bigi_Falk}
I. Bigi et al., Phys. Lett.~{\bf B323}, 408 (1994);\\
A. Falk, M.B. Wise, and I. Dunietz,  Phys. Rev.~{\bf D51}, 1183 (1995);\\
I. Dunietz et al.,  Eur. Phys. J.~{\bf C1}, 211 (1998);\\
H. Yamamoto, hep-ph/9912308.


\bibitem{Mellin}
V.A. Smirnov, Renormalization and Asymptotic Expansions,
Birkh\"auser Basel 1991;\\
E.E. Boos and A.I. Davydychev, Theor. Math. Phys. {\bf 89}
1052 (1992);\\
N.I. Usyukina, Theor. Math. Phys. {\bf 79} (1989) 385,
{\bf 22} 211 (1975);\\
A. Erdelyi (ed.), Higher Transcendental  Functions,
McGraw New York 1953.


\bibitem{Misiak97}
 K. Chetyrkin, M. Misiak, and M. M\"unz,
Phys. Lett.~{\bf B400}, 206 (1997);
Nucl. Phys.~{\bf B518}, 473 (1998);
Nucl. Phys.~{\bf B520}, 279 (1998).


\bibitem{Ciuchini}
 M. Ciuchini et al. Phys. Lett.~{\bf B334}, 137 (1994);  


\bibitem{Bagan}
E. Bagan et al., Nucl. Phys.~{\bf B432}, 3 (1994);
Phys. Lett.~{\bf B 342}, 362 (1995) [E:{\bf 374}, 363 (1996)]; 
E. Bagan et al., Phys. Lett.~{\bf B 351}, 546 (1995)

\bibitem{Japantalk}
A. Golutvin, plenary talk given at the XXXth International Conference
on High Energy Physics, Osaka, Japan, July 2000.


\end{thebibliography}
\end{document}